\documentclass[12pt]{article} 
\usepackage[T2A]{fontenc} 
\usepackage[utf8]{inputenc} 
\usepackage[russian, english]{babel} 
\usepackage{epsf}
\usepackage{wrapfig}
\usepackage{cyr}
\usepackage{pazh}
\tightenlines

\usepackage[T2A]{fontenc}
\usepackage[utf8]{inputenc}

\usepackage{wrapfig}
\usepackage{amssymb}
\usepackage{epsfig}
\usepackage{amsmath}
\usepackage{multirow}
\usepackage{caption}
\usepackage{subcaption}
\usepackage{soul}
\usepackage{float}
\usepackage[round]{natbib}

\DeclareMathOperator{\sign}{sign}
\sloppypar

\begin{document}
\baselineskip 21pt

\vspace{2cm}

\title{\bf 
DATA FROM FAST AND MEERKAT SURVEYS AS A TEST OF RADIO PULSAR PHYSICS}

\author{\bf \hspace{-1.3cm}\copyright\, 2025 г. \ \ 
F. A. Kniazev\affilmark{1,2},
A. Yu. Istomin\affilmark{1,2},
V. S. Beskin\affilmark{2,1\ast}
}

\affil{
{$^1$\it Moscow Institute of Physics and Technology, Dolgoprudny, Moscow region,Institutsky per. 9, 141700, Russia}, \\ 
{$^2$\it P.N.Lebedev Physical Institute, Leninsky prosp., 53, Moscow, 119991, Russia}\\ 
}


\sloppypar 
\vspace{2mm}
\noindent

The data from the FAST and MeerKAT surveys has significantly increased the number of radio pulsars for which the polarization characteristics of their mean profiles have been determined in detail. This has allowed us to confirm earlier conclusions both about the nature of propagation of two orthogonal modes in the pulsar magnetospheres and about the mechanism of particle production in neutron star polar regions and their evolutionary features. We can now say with even greater confidence that mean profiles formed by the O-mode are significantly wider than those formed by the X-mode. Moreover, the observations confirm the validity of the classical Ruderman-Sutherland vacuum model of particle generation, as well as the evolution of the inclination angles of the magnetic axis to the rotation one in the direction of $90^{\circ}$.

\noindent
{\it Key words:\/} radio pulsars, mean profiles, evolution


\vfill
\noindent\rule{8cm}{1pt}\\
{$^*$ E-mail: beskin@lpi.ru}


\clearpage

\baselineskip 21pt

\newpage

\section*{INTRODUCTION}

Despite the fact that over half a century has passed since the discovery of radio pulsars, there are still many key aspects related to their activity that we do not fully understand. For instance, there is no consensus on whether the angle between the magnetic axis and the rotation axis of a neutron star changes to small or large ($\sim 90^{\circ}$) values over time \citep{Lyne2015, Philippov_Kramer, Ablomasov}. Furthermore, the mechanism behind the coherent radio emission from pulsars remains a mystery.

Note that this was largely due to the fact that detailed catalogues containing fairly complete information on both the mean pulse emission profiles of pulsars and their polarization properties \citep{Taylor_Manchester, Rankin1983, hankinsrankin2010} did not exceed a few dozen pulsars. This is despite the fact that the number of known radio pulsars had long exceeded several thousand by the end of 2024: The ATNF catalogue \citep{ATNF} already contained 3,748 sources at that time. This, of course, significantly limited the possibilities of quantitative verification of the theories of evolution and radio emission of radio pulsars.

However, it has long been established that the two orthogonal modes observed in pulsar radio emission (see, for example, \cite{Stinebring1984}) are related to the ordinary (O) and extraordinary (X) modes, which naturally occur in magnetically active plasmas \citep{B&A}. At the same time, due to the refraction of the O-mode, corresponding mean intensity profiles should be noticeably wider than the mean profiles formed by the X-mode. For example, in the work of \cite{BGI88}, assuming a constant density of the outflowing plasma in the cross section, the following estimates were obtained for the characteristic widths of the mean profiles (periods $P$ hereafter in seconds):

{\footnote{The two variants for the ordinary mode correspond to the radiation region located near the neutron star or at the maximum possible distance from it.}}
\begin{eqnarray}
W_{\rm r,1}^{\rm O} & \approx & 7.8^{\circ} P^{-0.43} \nu_{\rm GHz}^{-0.14},
\label{Wr1} \\
W_{\rm r,2}^{\rm O}  & \approx & 10.8^{\circ} P^{-0.5} \nu_{\rm GHz}^{-0.29},
\label{Wr2} \\
W_{\rm r}^{\rm X}  & \approx & 3.6^{\circ} P^{-0.5} \nu_{\rm GHz}^{-0.5}. 
\label{Wr3}
\end{eqnarray}
However, confirmation of this seemingly simple fact has been challenging for many years, as there has been no method to determine with certainty which of the two orthogonal modes forms the mean profile for each individual pulsar.

The procedure was found after it was shown that the circular polarization of radiation emitted from the magnetosphere of a neutron star is not associated with a difficult-to-measure difference in the distributions of electrons and positrons (i.e. with non-diagonal terms of the dielectric tensor), but rather with a shear (rotation) of the external magnetic field in the direction of the line of sight. The direction of this shear, in turn, is connected to the observed variation in the position angle of linear polarization {\it p.a.} \citep{andrianovbeskin2010, wanglaihan2010, beskinphilippov2012}.
As a result, the signs of the derivative ${\rm d}p.a./{\rm d}\phi$ ($\phi$ --- pulse phase) and the Stokes parameter V, which determines the circular polarization, should be the same for the X-mode and opposite for the O-mode \citep{HBP}. However, until recently, only a few dozen pulsars had both the position angle track and the sign of the circular polarization known with good accuracy. This made it difficult to determine the difference in the properties of the ordinary and extraordinary modes with certainty.

The first catalogue, which contained sufficiently detailed polarization data for several hundred objects, was published in 2018 by Johnston and Kerr. At that time, 175 out of 600 pulsar have their position angle {\it p.a.} track and the sign of circular polarization V determined with good accuracy \citep{UFN}. As a result, it was shown that the widths of the mean profiles formed by the O-mode exceed those formed by the X-mode (see the Table ~\ref{table1} below). However, to confirm this final conclusion, it was necessary to increase the statistics of the observed sources significantly. This was achieved with the start of operation of the FAST and MeerKAT telescopes.

Indeed, the data published in recent years by the FAST \citep{FAST682} and MeerKAT \citep{MeerKAT_XI} radio telescopes have significantly increased the statistics of the polarization properties of the pulsar mean radio emission profiles. The first part of this paper is devoted to a detailed discussion of the results obtained. Based on the richer statistics, it will be shown that the widths of the mean profiles fully correspond to the predictions of the theory. 

Another problem which new data from FAST and MeerKAT can help to solve is the statistics of orthogonal interpulse radio pulsars. These are pulsars with a magnetic axis inclined at a near-90-degree angle to their axis of rotation, allowing us to observe radiation from both magnetic poles. We believe that these radio pulsars could serve as a test for neutron star evolution models \citep{UFN}. However, due to their relative rarity, their statistical properties have not yet been reliably determined.

First of all, we recall that the dynamics of a rotating neutron star is determined by the torque associated with the surface currents that close longitudinal (along the magnetic field) currents flowing in the magnetosphere. In the MHD model based on numerical simulations of the magnetosphere of a rotating neutron star \citep{spitkovsky2006, philippovtchekh2014}, the longitudinal currents flowing through the magnetosphere are determined from a global solution. For orthogonal rotators, they differ little from aligned rotators. As a result, in this model, the energy losses increase with increasing inclination angle of the magnetic axis relative to the rotation axis $\chi$, and the angle $\chi$ decreases over time.

In turn, in the BIG model \citep{bgi83, BGI93}, the longitudinal current is close to the so-called Goldreich-Julian current $j=c\rho_{\rm GJ}$, where the corresponding charge density
\begin{equation}
\rho_{\rm GJ} = -\frac{{\bf \Omega B}}{2 \pi c} \propto \cos\chi
\end{equation}

($\Omega = 2\pi/P$ is the angular frequency of the neutron star rotation) depends significantly on the inclination angle $\chi$. Therefore, for orthogonal pulsars, the total longitudinal current circulating in the magnetosphere is $(\Omega R/c)^{1/2}$ times less than for aligned rotators (i.e. it is 100 times smaller than for the ordinary pulsars). Accordingly, in this model, energy losses decrease as the inclination angle $\chi$ increases, and the inclination angle $\chi$ tends towards $90^{\circ}$.

Orthogonal pulsars are also interesting because they should have practically no current sheet outside the light cylinder \citep{spitkovsky2006, cont2012, philippovspitkovsky2014}. This feature is crucial in many modern models of the magnetosphere \citep{Philippov_Kramer}. However, the main features of orthogonal radio pulsars are not different from those of ordinary pulsars.

In addition, these two models predict significantly different electric potential drops in the polar regions of a neutron star, where particles are produced. In the case of the BGI model, based on the idea of a classical vacuum gap \citep{RS}, the accelerating potential of orthogonal pulsars should be significantly lower than that of aligned pulsars. However, numerical simulations using the particle-in-cells method, which is based on the assumption of a weak dependence of the longitudinal current on the inclination angle following from the currently actively developed MHD  \citep{Philippov_Kramer}model, do not show significant change in the accelerating potential \citep{Benachek2024}. Finally, the mere existence of orthogonal interpulse pulsars set limitations on the magnitude of the magnetic field required for the generation of secondary particles, that is, it imposes a restriction on the model of evolution, which determines this field. Therefore, the analysis of the observed number of pulsars can significantly help to determine the model of their evolution.

For the first time, the data from the MeerKAT and FAST observations have made it possible to obtain the relative number of orthogonal interpulse pulsars in a homogeneous sample. This confirms the conclusion obtained from smaller statistics and a heterogeneous sample that the number of such pulsars is at least 3\%. It has also been shown that the most of orthogonal pulsars have a period of about $P  \lesssim 0.2 \div 0.3 \text{ s}$. These results are discussed in detail in the final section of the article, as they challenge the currently accepted belief that the angle between the magnetic axis and the axis of rotation of a pulsar decreases over time. Other aspects of pulsar evolution are also addressed in that section.

\section*{WIDTHS OF THE RADIO PULSAR MEAN PROFILES}
\noindent

The pulsar emission modes were determined using two independent methods. First of all, the polarization characteristics of all 682 pulsars in the FAST catalogue and 1170 in the MeerKAT catalogue were manually examined. Accordingly, we have chosen 138 and 437 pulsars which, firstly, had mean profiles that were confidently associated with only one orthogonal mode (i.e. with a standard S-shaped positional angle track and no orthogonal mode jumps) and, secondly, for which the Stokes V parameter was also well-determined and did not change its sign over the pulse. Pulsars with similar signs of the derivative ${\rm d}p.a./{\rm d}\phi$ and Stokes V were assigned to the X-mode, while those with different signs were assigned to the O-mode. Thus, pulsars with single-humped mean profiles were labelled as Xs and Os, and those with double-humped profiles were labelled as Xd and Od. Recall that according to \cite{andrianovbeskin2010}, single-humped profiles are most likely to be associated with the X-mode, while double-humped profiles are mainly associated with the O-mode.

In addition, in order to eliminate human bias, a simple algorithm was developed to numerically determine the emission mode from digitized FAST{\footnote{\, Taken from http://zmtt.bao.ac.cn/psr-fast/}} and MeerKAT{\footnote{\,Taken from https://zenodo.org/records/7272361}} data. To do so, at each point in the profile where the radiation intensity exceeded 10\% of the maximum, the signs of circular polarization and derivative $d p.a. / d \phi$ were determined (for this purpose, the position angle track was pre-smoothed). After that, the quantity 
\begin{equation}
    \eta = \frac{1}{N} \sum_{I \geq 0.1 I_{max}} \sign{V} \cdot \sign{\frac{d p.a.}{d \phi}},
\end{equation}
was calculated, where N is the number of points which were used for summation. Thus, $-1 \leq \eta \leq 1$, $\eta = 1$ corresponds to the pure X-mode and $\eta = -1$ corresponds to the pure O-mode. Next, the classification can be performed if one selects a specific value $\eta_{cr}$. Pulsars with $\eta > \eta_{cr}$ will then be considered to emit primarily in the X-mode, while those with $\eta < -\eta_{cr}$ emit primarily in the O-mode. Here, the exact value of $\eta_{cr}$ determines the robustness of the algorithm used. 

In this study, the value of $\eta$ was chosen to be $\eta = 0.4$, as it allowed for the largest possible sample size to be obtained while keeping the mode determination error rate below 5\% (the comparison was made using the intersection of manually and automatically obtained samples). However, the sample obtained manually for pulsars from the MeerKAT catalogue was found to be significantly larger, mainly due to the pulsars with low signal-to-noise ratio of polarization profiles. For these pulsars, the algorithm was unable to unambiguously determine the radiation mode, while visual inspection in many cases allowed us to make this determination. Therefore, it was decided to test the main hypothesis regarding the difference in mean profile widths using samples obtained via both methods, but to provide detailed results only for the manually obtained samples.

Further, the width of the mean profiles was determined through direct computation of $W_{10}$ based on FAST and MeerKAT data. We have chosen to use the value of $W_{10}$, as for the pulsars with double-humped profiles, for which the intensity of one of the humps is less than 50\%, $W_{50}$ quantity poorly represents the actual profile width. Finally, this paper considers the distribution of $W_{10}P^{1/2}$, which, according to relations (\ref{Wr1})--(\ref{Wr3}), allows us to eliminate the dependence on period $P$ with great accuracy. No frequency correction was performed, as the frequencies used (1250 MHz for FAST and 1284 MHz for MeerKAT) are practically identical to each other and close to the 1 GHz normalization used for the estimates (\ref{Wr1})--(\ref{Wr3}).

\begin{table}
\caption{The observed number of pulsars and median widths of their mean profiles $W(^\circ) P^{1/2}$ for each of the modes for single-humped (S) and double-humped(D) mean profiles.} 
\vspace{0.3cm}
\centering
\begin{tabular}{|c|c|cccc|cc|}
\hline
              &  $N$   &  Xs  &   Xd & Os  & Od  & X$_{\rm tot}$ & O$_{\rm tot}$\\
\hline
FAST &  39+19+19+37
  &  $12.4_{-2.6}^{+2.4}$    &  $12.7_{-3.3}^{+5.7}$   &  $12.6_{-1.5}^{+6.1}$  & $16.0_{-2.3}^{+2.7}$ & $12.5_{-2.2}^{+2.3}$ & $15.4_{-2.2}^{+2.5}$\\  
\hline
MeerKAT & 178+36+93+114
  &   $10.2_{-0.7}^{+1.4}$   &   $11.6_{-1.6}^{+2.9}$  & $12.9_{-1.1}^{+1.9}$  & $16.8_{-1.5}^{+2.0}$ & $10.6_{-0.9}^{+1.1}$  & $15.3_{-1.8}^{+0.8}$\\  
\hline
\end{tabular}
\label{table1}
\end{table}

\begin{figure}[H]
\centering
\begin{subfigure}{.49\textwidth}
  \centering
  \includegraphics[width=.8\linewidth]{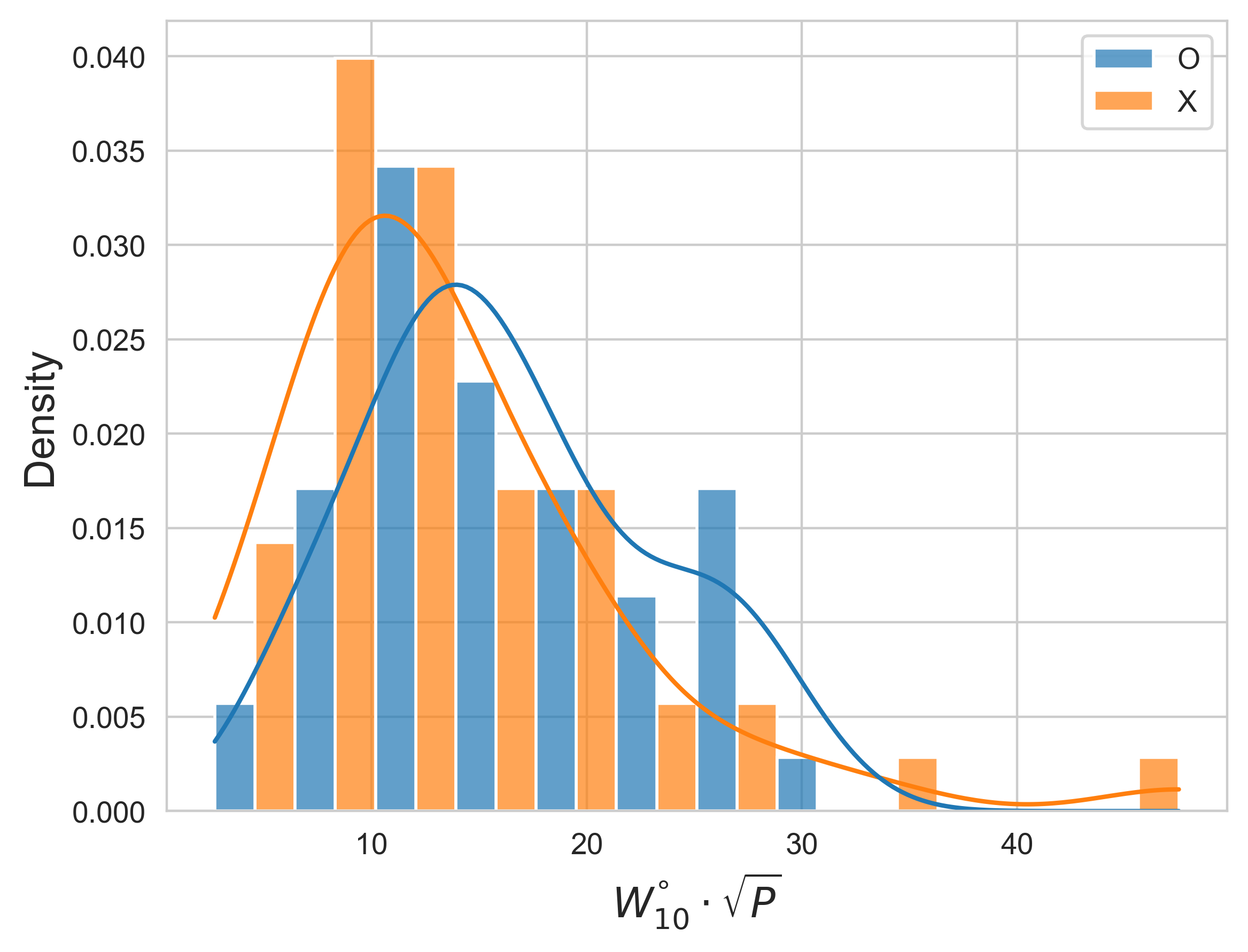}
  \caption{FAST}
  \label{fig:sub1}
\end{subfigure}%
\begin{subfigure}{.49\textwidth}
  \centering
  \includegraphics[width=.8\linewidth]{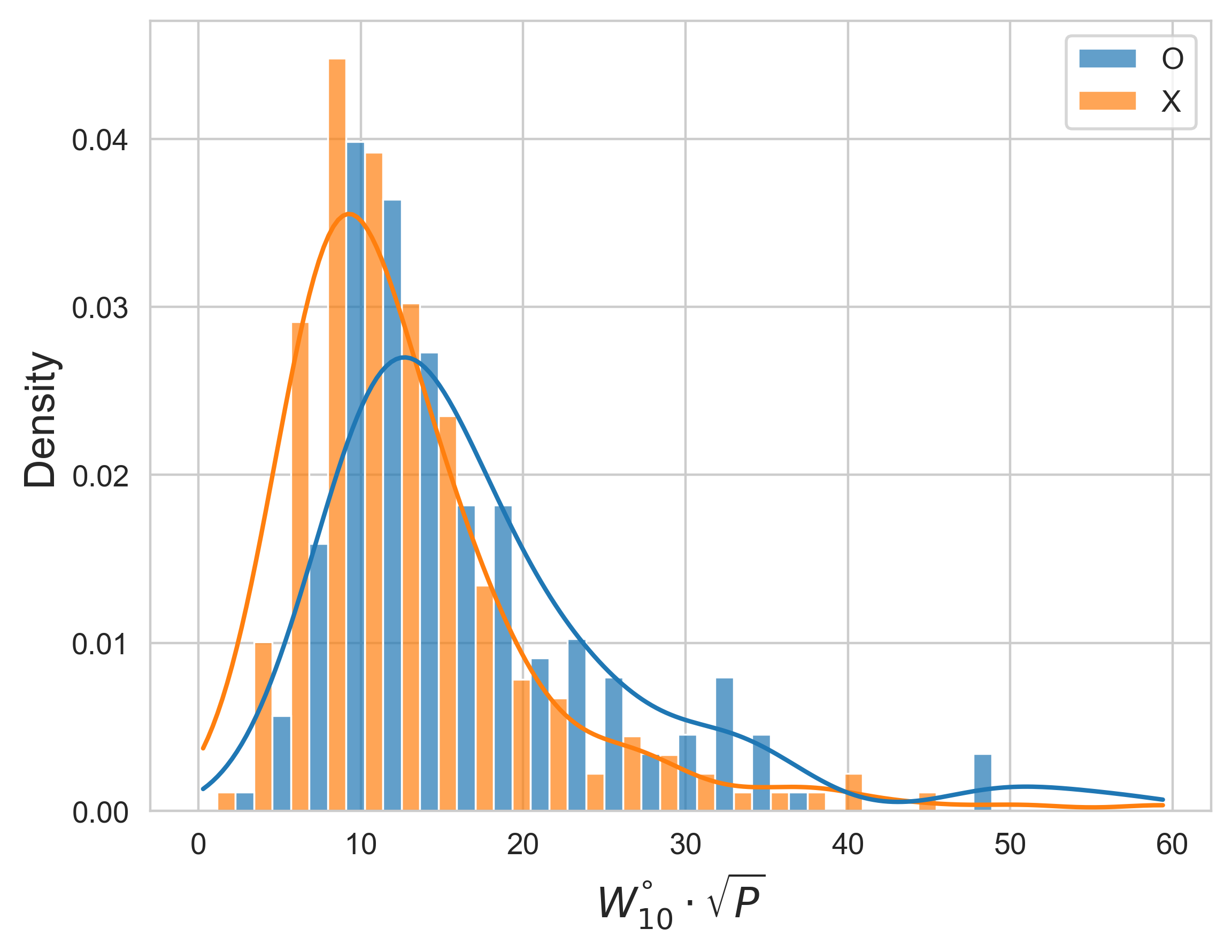}
  \caption{MeerKAT}
  \label{fig:sub2}
\end{subfigure}
\caption{The distribution of the widths of pulsar profiles with a certain mode (normalization per unit is chosen). The solid lines correspond to the kernel density estimates of these distributions.}
\label{fig1}
\end{figure}

\begin{table}[H]
\centering
\caption{The p-value for statistical tests used to compare $X$ and $O$ mode samples.} 
\begin{tabular}{|l|llll|}
\hline
\multirow{2}{*}{p-value} & \multicolumn{4}{l|}{The mode detection method}                                                   \\ \cline{2-5} 
                              & \multicolumn{2}{l|}{Manual}                              & \multicolumn{2}{l|}{Automated} \\ \hline
Statistical test & \multicolumn{1}{l|}{FAST} & \multicolumn{1}{l|}{MeerKAT} & \multicolumn{1}{l|}{FAST} & MeerKAT \\ \hline
Anderson–Darling test       & \multicolumn{1}{l|}{0.03} & \multicolumn{1}{l|}{0.001}   & \multicolumn{1}{l|}{0.04} & 0.005   \\ \hline
permutation test           & \multicolumn{1}{l|}{0.02} & \multicolumn{1}{l|}{0.0001}  & \multicolumn{1}{l|}{0.02} & 0.009   \\ \hline

\end{tabular}
\label{table:tests}
\end{table}

The corresponding distributions of the obtained profiles are shown in Figure \ref{fig1}. Table~\ref{table1} provides median values of $W_{10}P^{1/2}$ for each mode, according to the classification defined earlier, along with 95\% confidence intervals. As we can see, the median widths of mean profiles formed by the O-mode are indeed larger than for those formed by the X-mode. Furthermore, single-humped profiles (where the line of sight passes laterally through the hollow-cone beam pattern) are always narrower than double-humped profiles (central passage). The most statistically significant difference was found between the widths of double-humped profiles for the X and O modes. Indeed, purely from geometric considerations, the beam broadening effect for the O-mode is expected to be most pronounced for central passage of the line-of-sight, i.e. for the double-humped profiles.

Analysing the combined statistics of single- and double-humped profiles, we found that the corresponding confidence intervals overlap for the FAST catalogue sample but do not overlap for the MeerKAT sample, which may be caused by the smaller size of the former. Therefore, to rigorously test the hypothesis, appropriate statistical tests were performed. To assess differences between the distributions, the Anderson-Darling test and a permutation test (at a 95\% significance level) were applied. For the latter, the difference in medians between samples was used as the test statistic. As shown in Table \ref{table:tests}, for both catalogues (MeerKAT and FAST) and both mode classification methods, the null hypothesis of the same distribution functions for the X and O modes is rejected by all tests applied. Thus, we conclude that the corresponding distribution functions have statistically significant differences.

However, it is important to note that direct comparison of the obtained median values with theoretical predictions is significantly challenging. While equations \eqref{Wr1}, \eqref{Wr2}, and \eqref{Wr3} effectively define the full beam width, the observed distributions of profile widths depend on the (a priori unknown) distributions of pulsar viewing and inclination angles. Another critical factor is that the radiation intensity distribution across the beam pattern for real pulsars is highly non-uniform, whereas equations \eqref{Wr1}, \eqref{Wr2}, and \eqref{Wr3} were derived in an assumption of the uniform distribution. Therefore, construction of a theoretical model required to analyse the observed distributions demands a dedicated study, which is beyond the scope of this paper.

Nevertheless, the conclusion about the broadening of the O-mode intensity profiles remains valid, and theoretical predictions regarding the dependence of morphological properties of the mean intensity profiles on emission mode correspond to the latest observational data. In addition, the method for determining the dominant emission mode in radio pulsars has been indirectly confirmed, as it allowed to successfully separate pulsars into two samples with distinctly different properties. Thus, the theory of radio wave propagation in pulsar magnetospheres has received yet another confident confirmation.

\section*{STATISTICS OF ORTHOGONAL INTERPULSE PULSARS AS A TEST OF THE EVOLUTION MODEL}
\subsection*{TOTAL NUMBER OF ORTHOGONAL INTERPULSE PULSARS}

First of all, we note that the MHD model \citep{philippovtchekh2014} , where the inclination angle $\chi$ decreases over time, and the BGI model \citep{bgi83, BGI93}, which predicts that $\chi$ tends toward $90^{\circ}$, should yield different predictions for the observed fraction of orthogonal interpulse pulsars. As was shown in \cite{Novoselov}, the MHD model predicts that the relative number of orthogonal interpulse pulsars in the period range $0.033$ s $< P < 0.5$ s should not exceed 1\% of the total number of pulsars in this range, whereas the BGI model predicts values of around (2.5–5.5)\%. At the same time, observational data provided in that work indicate that the relative fraction of orthogonal interpulse pulsars in this range is about (1.8–2.6)\%, which, in general, favours the BGI model.  

However, this conclusion depends on the secondary plasma generation model, as the density distribution of this plasma significantly affects the ability to observe both magnetic poles. Although the model used in \cite{Novoselov} relied on simplified assumptions (e.g., the acceleration gap height and potential were estimated using algebraic relations from the one-dimensional Ruderman-Sutherland vacuum gap model, which is inapplicable for orthogonal rotators), a recent more precise calculation \citep{Istomin2025}, where the accelerating potential was obtained self-consistently, yielded nearly identical results.  

It is also important to note that the observational data analysed in \cite{Novoselov} were compiled from surveys with varying sensitivities. Furthermore, the result itself required refinement, as the obtained percentage was close to the boundary of the predicted range and only slightly exceeded the MHD model’s $\sim 1\%$ prediction. Here, the homogeneous datasets from FAST and MeerKAT proved invaluable, enabling the first robust determination of the relative fraction of orthogonal interpulse pulsars.  

For the FAST data, we used Fig. A6 from \cite{FAST682}, which contains 27 mean profiles of interpulse pulsars. We have chosen 26 objects among them based on a phase separation of $180^{\circ}$ between the main pulse and interpulse --- a criterion widely accepted for identifying orthogonal interpulse pulsars (e.g., \cite{WJ2008}).  

We emphasise that the number of orthogonal interpulse pulsars derived from Fig. A6 should be considered as a lower limit. Only 10\% of the pulsars in this catalogue have an interpulse-to-main-pulse intensity ratio below 0.01, compared to over 20\% in the heterogeneous catalogue used by \cite{Novoselov}. Thus, some interpulse pulsars may have been excluded from Fig. A6. Conversely, it is possible that some pulsars with near-$180^{\circ}$ phase separation are aligned rotators rather than orthogonal ones, leading to an overestimation. Taking both factors into account, we adopt 26 as the most reliable estimate.  

For the MeerKAT data \citep{MeerKAT_VI}, orthogonal interpulse pulsars were identified as sources with unambiguous separation between the main pulse and interpulse and a phase separation of $\sim180^{\circ}$.  

\begin{table}
\caption{The observed absolute and relative numbers of orthogonal interpulse pulsars for various surveys are presented. The values of the total number of observed pulsars within these ranges of periods are also provided.}
\vspace{0.3cm}
\centering
\begin{tabular}{|c|c|c|c|c|}
\hline
 &  $P < 0.033 \, {\rm c}$   &   $0.033 \, {\rm c} < P < 0.5 \,{\rm c}$ & $0.5 \,{\rm c} < P < 1 \,{\rm c}$  & $P > 1 \,{\rm c} $ \\
\hline
Novoselov et al  &  $-$   &  (18 $\div$ 26)/968 & (3 $\div$ 5)/725  & (0 $\div$ 1)/694 \\  
  &   $-$   &   (1.8$\div$2.6)\%     &  (0.4$\div$ 0.7)\% &   (0.0$\div$0.1)\% \\  
\hline
FAST    &   8/73 &  14 /233    &  3/177&  1/198 \\ 
  &  11.0\%    &   6.0\%     &  1.7\%  & 0.5\% \\  
\hline
MeerKAT  &  0/1  &   25/590   &  4/414  & 0/265 \\ 
  &  0\%   &  4.2\%     &  1.0\%  &  0\% \\  
\hline 
BSA  &  0/0   &   (1 $\div$ 2)/28    &  (0 $\div$ 1)/27 & 1/41\\ 
  &  0\%   &   (3$\div$7)\%     & (0$\div$3)\%   &  2\% \\  
\hline 
\end{tabular}
\label{table3}
\end{table}

As shown in Table~\ref{table3}, the data from the FAST and MeerKAT observatories are in good agreement with each other and with the results of \cite{Noutsos2015}. In all cases, the relative number of orthogonal interpulse pulsars in the period range $0.033$ s $< P < 0.5$ s significantly exceeds the number of pulsars with periods $P > 0.5$ s. Despite limited statistics, we also included data from the BSA PRAO observatory \citep{Toporov} in Table~\ref{table3}, which, as can be seen, do not contradict the results from other observatories.  

The discrepancy between the FAST and MeerKAT data presented here and those in \cite{Toporov} arises because we present the percentage of interpulse pulsars relative to the number of sources in each period range, rather than the total population of observed radio pulsars. As one can see, the PRAO data are also consistent with the data from other observatories. All combined, these results strongly favour the evolution of the inclination angle $\chi$ towards $90^{\circ}$.
 
\subsection*{DISTRIBUTION OF THE NUMBER OF ORTHOGONAL PULSARS BY PERIOD}

Another important difference between the MHD and BGI models is that for the orthogonal pulsars, they provide significantly different values for the electric potential drop in the region above the poles of the neutron star, where particles are produced. In the BGI model, this maximum possible potential drop near the polar caps is determined using relation for the vacuum gap, given in the work of \cite{RS}
\begin{equation}
\psi_{\rm max} \approx 2 \pi \rho_{\rm GJ}R_{0}^2, 
\label{psi}
\end{equation}
where $R_{0}$ is the radius of the polar cap. As a result, we obtain a restriction for the generation of secondary particles near the polar caps \citep{bgi83}
\begin{equation}
\cos\chi > k \, P^{15/7}\,  B_{12}^{-8/7},
\label{dl}
\end{equation}
where, as usual, $B_{12} = B_{0}/(10^{12}$ G), and $k \approx 1$.

The death line defined in this way is fundamental for the BGI model. In turn, in the MHD model, the potential drop is determined by the longitudinal electric current flowing in the magnetosphere. Therefore, the potential drop in the region of secondary plasma generation for orthogonal rotators differs little from the case of an aligned rotator (see, for example, \cite{Benachek2024} {\footnote{In this work, for an orthogonal rotator, the longitudinal electric field is $\sim 10\%$ of the field on the axis for an aligned rotator, whereas for the vacuum gap in the case \mbox{$P = 0.25$ s} considered in this work, this ratio cannot exceed 1\% \citep{PaperII}.}}. Here, we will show that the new data from the FAST and MeerKAT surveys help to clarify this issue.

First of all, we note that the work of \cite{Novoselov} has already demonstrated that the distribution of pulsars on the ($\chi$--$P\,B^{8/15}$) plane exactly corresponds to the relation (\ref{dl}) for $k_{\rm MHD} = 1.0\pm 0.1$ and $k_{\rm BGI} = 0.5\pm 0.1$. Thus, it is now difficult to doubt the validity of the death line defined by equation (\ref{dl}).

Furthermore, the data from the FAST and MeerKAT surveys confirmed the small number of orthogonal interpulse pulsars with periods $P > 0.5$ s. This property is a direct consequence of the death line relation (\ref{dl}), since, as already noted, within this model, the accelerating potential in the limiting case of a vacuum gap for an orthogonal rotator should be smaller by a factor of $(\Omega R/c)^{1/2}$, i.e. tens or even hundreds of times less than in the aligned case. As a result, substituting the characteristic value of $(\Omega R/c)^{1/2}$ for $\cos\chi$ into expression (\ref{dl}), we obtain  
\begin{equation}  
P < 0.2 \, B_{12}^{16/37} {\rm s}.  
\label{P}  
\end{equation}  
Consequently, for average magnetic field values of around $B_{12} \approx 1$, the peak in the distribution of orthogonal interpulse pulsars should fall within the period range $P \sim (0.1-0.3)$ s \citep{Novoselov}.  

\begin{table}
\caption{The observed number of orthogonal interpulse pulsars for various surveys in the periods range of $0.03$ s$< P < 0.5$s as a percentage.}
\vspace{0.3cm}
\centering
\begin{tabular}{|c|c|c|c|c|c|}
\hline
$P$ (с) &  0.03--0.1   &   0.1--0.2  &  0.2--0.3 & 0.3--0.4 & 0.4--0.5 \\
\hline
BGI  &  0.0   &  1.2   & 0.6  & 0.3 & 0.2 \\  
\hline
Novoselov et al.  & 0.4 $\div$ 0.6    &  1.0 $\div$ 1.9   & 1.4 $\div$ 1.6    & 0.6 $\div$ 0.8 &   0.2 $\div$ 0.4 \\  
\hline
FAST    &   0.4 &  2.1   &  1.7 &  1.3 &  0.4 \\ 
\hline
MeerKAT  &  0.3   &   1.2     &  1.5  & 1.0 & 0.3 \\ 
\hline 
\end{tabular}
\label{table4}
\end{table}

Table ~\ref{table4} shows the number of observed orthogonal interpulse pulsars within the period range of $0.03$ s $< P < 0.5$ s both for the new FAST and MeerKAT surveys, as well as the compilation data from the article of \cite{Novoselov}. In addition, in the top row predictions of the BGI model are provided, which were determined in that same work based on an analysis of the kinetic equation where inequality (\ref{dl}) was used. As we can see, the data from the FAST and MeerKAT surveys are consistent with the BGI model, which predicts, according to equation (\ref{P}), a maximum in the distribution of orthogonal interpulse pulsars near $0.2 - 0.3$ seconds. At the same time, they are inconsistent with the MHD model, in which, as we have already noted, the dependence of the death line (\ref{dl}) on the inclination angle is insignificant and, therefore, the number of orthogonal interpulse pulsars with the periods of around $P \sim 1$ s should be comparable to the number of such pulsars in the range $0.03$ c $< P < 0.5$. 

It can also be noted that according to the results of the FAST observatory, there is an even larger relative number of orthogonal interpulse pulsars with the periods below 0.033 s. However, as is known, the evolution of millisecond radio pulsars differs significantly from that of the ordinary pulsars, and therefore the conclusions of Novoselov et al. \cite{Novoselov} do not apply to them.

\section*{DISCUSSION AND CONCLUSION}
\noindent

Thus, we can confidently conclude that the FAST and MeerKAT surveys confirm earlier findings (based on significantly smaller statistical samples) that,  first, the observed widths of mean profiles of radio pulsars generated by the X-mode are notably smaller than those generated by the O-mode. Secondly, the statistics of orthogonal radio pulsars is better explained by the BGI evolutionary model compared to the MHD model.The first conclusion is in full agreement with the predictions of the currently accepted theory of wave propagation in neutron star magnetospheres (see, e.g., \cite{P&L1, L&GS, Noutsos2015} ) and, therefore, does not require additional discussion.  
As for the second conclusion, it, as we see, contradicts the currently accepted viewpoint on the evolutionary direction of the inclination angle of the magnetic axis relative to the rotation one.
 
It should be recalled that direct measurement of the evolutionary direction of the inclination angle (the sign of the derivative $\dot \chi$) is far beyond the precision of modern telescopes. Therefore, in order to determine this quantity, various indirect procedures based on statistical distributions should be used \citep{T&M, Lyne2015}. For example, using the observational estimation of the pulsar distribution function $f(\chi, P)$ over $\chi$ and period $P$, which should, in addition, predict the number of observed interpulse pulsars.  

For instance, it was established long ago that the averaged inclination angle <$\chi(P)$> decreases with increasing period $P$ \citep{T&M}. Correspondingly, the averaged widths of mean profiles <$W_{r}(P)$> also increase with the period \footnote{For an intrinsic beam width $W_{0}$, the observed mean profile width depends on the angle $\chi$ as $W_{r} \approx W_{0}/\sin\chi$.} \citep{RankinIV, Faucher, WJ2008b, Gullon2014}. For many years, according to numerous authors \citep{Young2010, MaciesiakW50, MeerKAT_XI}, these results have been considered as a direct evidence that the angle $\chi$ decreases over time for each individual pulsar.  

However, this behaviour of the averaged inclination angle <$\chi(P)$> has another explanation \citep{bgi83, BGI93}. It is based on the existence of the death line (\ref{dl}), i.e. the assumption that the equation (\ref{psi}) determines the maximum possible potential drop in the region of secondary plasma generation. Consequently, at sufficiently large periods $P$, the generation of secondary electron-positron plasma becomes possible only at small inclination angles $\chi$, resulting in the statistically averaged inclination angle decreasing with increasing $P$ regardless of the evolutionary direction. As was shown in \cite{bgi83} through analysis of the kinetic equation describing the entire radio pulsar population, the averaged values  <$\chi(P)$> must decrease with $P$ regardless of the evolutionary direction of the individual pulsars. Thus, it can be confidently stated that the BGI model, in which the death line (\ref{dl}) plays a key role, not only does not contradict observational data but is also confirmed by the latest FAST and MeerKAT surveys.  

Therefore, we would like to once again draw attention to orthogonal interpulse pulsars, which, in our opinion, through their very existence and the details of their statistical properties, could help to clarify many unresolved questions related to radio pulsars. This will undoubtedly be facilitated by the significant increase in statistics enabled due to the start of operation of the FAST and MeerKAT telescopes.  

\section*{ACKNOWLEDGMENTS}
The authors would like to thank A.Spitkovsky, A.A.Philippov and A.V.Chernoglazov for a useful discussion. The work was carried out with the support of the Russian Science Foundation (project 24-22-00120).

\bibliographystyle{plainnat}
\bibliography{references}

\begin{thebibliography}{41}
\providecommand{\natexlab}[1]{#1}
\providecommand{\url}[1]{\texttt{#1}}
\expandafter\ifx\csname urlstyle\endcsname\relax
  \providecommand{\doi}[1]{doi: #1}\else
  \providecommand{\doi}{doi: \begingroup \urlstyle{rm}\Url}\fi

\bibitem[{Abolmasov} et~al.(2024){Abolmasov}, {Biryukov}, and {Popov}]{Ablomasov}
Pavel {Abolmasov}, Anton {Biryukov}, and Sergei~B. {Popov}.
\newblock {Spin Evolution of Neutron Stars}.
\newblock \emph{Galaxies}, 12\penalty0 (1):\penalty0 7, February 2024.
\newblock \doi{10.3390/galaxies12010007}.

\bibitem[{Andrianov} and {Beskin}(2010)]{andrianovbeskin2010}
A.~S. {Andrianov} and V.~S. {Beskin}.
\newblock {Limiting polarization effect --- a key link in investigating the mean profiles of radio pulsars}.
\newblock \emph{Astronomy Letters}, 36:\penalty0 248--259, April 2010.
\newblock \doi{10.1134/S1063773710040031}.

\bibitem[{Barnard} and {Arons}(1986)]{B&A}
J.~J. {Barnard} and J.~{Arons}.
\newblock {Wave propagation in pulsar magnetospheres - Refraction of rays in the open flux zone}.
\newblock \emph{\apj}, 302:\penalty0 138--162, March 1986.
\newblock \doi{10.1086/163979}.

\bibitem[{Benáček, Jan} et~al.(2024){Benáček, Jan}, {Timokhin, Andrey}, {Muñoz, Patricio A.}, {Jessner, Axel}, {Rievajová, Tatiana}, {Pohl, Martin}, and {Büchner, Jörg}]{Benachek2024}
{Benáček, Jan}, {Timokhin, Andrey}, {Muñoz, Patricio A.}, {Jessner, Axel}, {Rievajová, Tatiana}, {Pohl, Martin}, and {Büchner, Jörg}.
\newblock Poynting flux transport channels formed in polar cap regions of neutron star magnetospheres.
\newblock \emph{A\&A}, 691:\penalty0 A137, 2024.
\newblock \doi{10.1051/0004-6361/202450949}.
\newblock URL \url{https://doi.org/10.1051/0004-6361/202450949}.

\bibitem[{Beskin}(2018)]{UFN}
V.~S. {Beskin}.
\newblock {Radio pulsars: already fifty years!}
\newblock \emph{Physics Uspekhi}, 61\penalty0 (4):\penalty0 353--380, April 2018.
\newblock \doi{10.3367/UFNe.2017.10.038216}.

\bibitem[{Beskin} and {Istomin}(2022)]{PaperII}
V.~S. {Beskin} and A.~Yu {Istomin}.
\newblock {Pulsar death line revisited - II. 'The death valley'}.
\newblock \emph{\mnras}, 516\penalty0 (4):\penalty0 5084--5091, November 2022.
\newblock \doi{10.1093/mnras/stac2423}.

\bibitem[{Beskin} and {Philippov}(2012)]{beskinphilippov2012}
V.~S. {Beskin} and A.~A. {Philippov}.
\newblock {On the mean profiles of radio pulsars - I. Theory of propagation effects}.
\newblock \emph{\mnras}, 425:\penalty0 814--840, September 2012.
\newblock \doi{10.1111/j.1365-2966.2012.20988.x}.

\bibitem[{Beskin} et~al.(1983){Beskin}, {Gurevich}, and {Istomin}]{bgi83}
V.~S. {Beskin}, A.~V. {Gurevich}, and I.~N. {Istomin}.
\newblock {The electrodynamics of a pulsar magnetosphere}.
\newblock \emph{Sov. Phys. JETP}, 58:\penalty0 235--253, August 1983.

\bibitem[{Beskin} et~al.(1988){Beskin}, {Gurevich}, and {Istomin}]{BGI88}
V.~S. {Beskin}, A.~V. {Gurevich}, and I.~N. {Istomin}.
\newblock {Theory of the radio emission of pulsars}.
\newblock \emph{\apss}, 146:\penalty0 205--281, July 1988.
\newblock \doi{10.1007/BF00637577}.

\bibitem[{Beskin} et~al.(1993){Beskin}, {Gurevich}, and {Istomin}]{BGI93}
V.~S. {Beskin}, A.~V. {Gurevich}, and Ya.~N. {Istomin}.
\newblock \emph{{Physics of the Pulsar Magnetosphere}}.
\newblock {Cambridge University Press}, Cambridge, August 1993.

\bibitem[{Faucher-Gigu{\' e}re} and {Kaspi}(2006)]{Faucher}
C.-A. {Faucher-Gigu{\' e}re} and V.~M. {Kaspi}.
\newblock {Birth and Evolution of Isolated Radio Pulsars}.
\newblock \emph{\apj}, 643:\penalty0 2401--2414, May 2006.
\newblock \doi{10.1086/501516}.

\bibitem[{Gull{\'o}n} et~al.(2014){Gull{\'o}n}, {Miralles}, {Vigan{\`o}}, and {Pons}]{Gullon2014}
Miguel {Gull{\'o}n}, Juan~A. {Miralles}, Daniele {Vigan{\`o}}, and Jos{\'e}~A. {Pons}.
\newblock {Population synthesis of isolated neutron stars with magneto-rotational evolution}.
\newblock \emph{\mnras}, 443\penalty0 (3):\penalty0 1891--1899, September 2014.
\newblock \doi{10.1093/mnras/stu1253}.

\bibitem[{Hakobyan} et~al.(2017){Hakobyan}, {Beskin}, and {Philippov}]{HBP}
H.~L. {Hakobyan}, V.~S. {Beskin}, and A.~A. {Philippov}.
\newblock {On the mean profiles of radio pulsars II: Reconstruction of complex pulsar light-curves and other new propagation effects}.
\newblock \emph{\mnras}, 469:\penalty0 2704--2719, March 2017.
\newblock \doi{10.1093/mnras/stx1025}.

\bibitem[{Hankins} and {Rankin}(2010)]{hankinsrankin2010}
T.~H. {Hankins} and J.~M. {Rankin}.
\newblock {Arecibo Multi-Frequency Time-Aligned Pulsar Average-Profile and Polarization Database}.
\newblock \emph{\aj}, 139:\penalty0 168--175, January 2010.
\newblock \doi{10.1088/0004-6256/139/1/168}.

\bibitem[{Istomin} et~al.(2024){Istomin}, {Kniazev}, and {Beskin}]{Istomin2025}
A.~Yu. {Istomin}, F.~A. {Kniazev}, and V.~S. {Beskin}.
\newblock {Acceleration Potential and Density Profile of Secondary Plasma in the Magnetosphere of Orthogonal Pulsars}.
\newblock \emph{Astronomy Reports}, 68\penalty0 (12):\penalty0 1271--1285, December 2024.
\newblock \doi{10.1134/S1063772925701264}.

\bibitem[{Johnston} et~al.(2023){Johnston}, {Kramer}, {Karastergiou}, {Keith}, {Oswald}, {Parthasarathy}, and {Weltevrede}]{MeerKAT_XI}
S.~{Johnston}, M.~{Kramer}, A.~{Karastergiou}, M.~J. {Keith}, L.~S. {Oswald}, A.~{Parthasarathy}, and P.~{Weltevrede}.
\newblock {The Thousand-Pulsar-Array programme on MeerKAT - XI. Application of the rotating vector model}.
\newblock \emph{\mnras}, 520\penalty0 (4):\penalty0 4801--4814, April 2023.
\newblock \doi{10.1093/mnras/stac3636}.

\bibitem[{Kalapotharakos} et~al.(2012){Kalapotharakos}, {Contopoulos}, and {Kazanas}]{cont2012}
C.~{Kalapotharakos}, I.~{Contopoulos}, and D.~{Kazanas}.
\newblock {The extended pulsar magnetosphere}.
\newblock \emph{\mnras}, 420:\penalty0 2793--2798, March 2012.
\newblock \doi{10.1111/j.1365-2966.2011.19884.x}.

\bibitem[{Lyne} and {Graham-Smith}(2012)]{L&GS}
A.~{Lyne} and F.~{Graham-Smith}.
\newblock \emph{{Pulsar Astronomy}}.
\newblock {Cambridge University Press}, Cambridge, March 2012.

\bibitem[{Lyne} et~al.(2015){Lyne}, {Jordan}, {Graham-Smith}, {Espinoza}, {Stappers}, and {Weltevrede}]{Lyne2015}
A.~G. {Lyne}, C.~A. {Jordan}, F.~{Graham-Smith}, C.~M. {Espinoza}, B.~W. {Stappers}, and P.~{Weltevrede}.
\newblock {45 years of rotation of the Crab pulsar}.
\newblock \emph{\mnras}, 446\penalty0 (1):\penalty0 857--864, January 2015.
\newblock \doi{10.1093/mnras/stu2118}.

\bibitem[{Lyubarskii} and {Petrova}(1998)]{P&L1}
Y.~E. {Lyubarskii} and S.~A. {Petrova}.
\newblock {Refraction of radio waves in pulsar magnetospheres}.
\newblock \emph{\aap}, 333:\penalty0 181--187, May 1998.

\bibitem[Maciesiak et~al.(2012)Maciesiak, Gil, and Melikidze]{MaciesiakW50}
Krzysztof Maciesiak, Janusz Gil, and Giorgi Melikidze.
\newblock {On the pulse-width statistics in radio pulsars – III. Importance of the conal profile components}.
\newblock \emph{\mnras}, 424\penalty0 (3):\penalty0 1762--1773, 08 2012.
\newblock ISSN 0035-8711.
\newblock \doi{10.1111/j.1365-2966.2012.21246.x}.
\newblock URL \url{https://doi.org/10.1111/j.1365-2966.2012.21246.x}.

\bibitem[Manchester et~al.(2005)Manchester, Hobbs, Teoh, and Hobbs]{ATNF}
R.~N. Manchester, G.~B. Hobbs, A.~Teoh, and M.~Hobbs.
\newblock The australia telescope national facility pulsar catalogue.
\newblock \emph{\apj}, 129\penalty0 (4):\penalty0 1993--2006, apr 2005.
\newblock \doi{10.1086/428488}.
\newblock URL \url{https://doi.org/10.1086/428488}.

\bibitem[{Noutsos} et~al.(2015){Noutsos}, {Sobey}, {Kondratiev}, {Weltevrede}, {Verbiest}, {Karastergiou}, {Kramer}, {Kuniyoshi}, {Alexov}, {Breton}, {Bilous}, {Cooper}, {Falcke}, {Grie{\ss}meier}, {Hassall}, {Hessels}, {Keane}, {Os{\l}owski}, {Pilia}, {Serylak}, {Stappers}, {ter Veen}, {van Leeuwen}, {Zagkouris}, {Anderson}, {B{\"a}hren}, {Bell}, {Broderick}, {Carbone}, {Cendes}, {Coenen}, {Corbel}, {Eisl{\"o}ffel}, {Fender}, {Garsden}, {Jonker}, {Law}, {Markoff}, {Masters}, {Miller-Jones}, {Molenaar}, {Osten}, {Pietka}, {Rol}, {Rowlinson}, {Scheers}, {Spreeuw}, {Staley}, {Stewart}, {Swinbank}, {Wijers}, {Wijnands}, {Wise}, {Zarka}, and {van der Horst}]{Noutsos2015}
A.~{Noutsos}, C.~{Sobey}, V.~I. {Kondratiev}, P.~{Weltevrede}, J.~P.~W. {Verbiest}, A.~{Karastergiou}, M.~{Kramer}, M.~{Kuniyoshi}, A.~{Alexov}, R.~P. {Breton}, A.~V. {Bilous}, S.~{Cooper}, H.~{Falcke}, J.~M. {Grie{\ss}meier}, T.~E. {Hassall}, J.~W.~T. {Hessels}, E.~F. {Keane}, S.~{Os{\l}owski}, M.~{Pilia}, M.~{Serylak}, B.~W. {Stappers}, S.~{ter Veen}, J.~{van Leeuwen}, K.~{Zagkouris}, K.~{Anderson}, L.~{B{\"a}hren}, M.~{Bell}, J.~{Broderick}, D.~{Carbone}, Y.~{Cendes}, T.~{Coenen}, S.~{Corbel}, J.~{Eisl{\"o}ffel}, R.~{Fender}, H.~{Garsden}, P.~{Jonker}, C.~{Law}, S.~{Markoff}, J.~{Masters}, J.~{Miller-Jones}, G.~{Molenaar}, R.~{Osten}, M.~{Pietka}, E.~{Rol}, A.~{Rowlinson}, B.~{Scheers}, H.~{Spreeuw}, T.~{Staley}, A.~{Stewart}, J.~{Swinbank}, R.~{Wijers}, R.~{Wijnands}, M.~{Wise}, P.~{Zarka}, and A.~{van der Horst}.
\newblock {Pulsar polarisation below 200 MHz: Average profiles and propagation effects}.
\newblock \emph{\aap}, 576:\penalty0 A62, April 2015.
\newblock \doi{10.1051/0004-6361/201425186}.

\bibitem[{Novoselov} et~al.(2020){Novoselov}, {Beskin}, {Galishnikova}, {Rashkovetskyi}, and {Biryukov}]{Novoselov}
E.~M. {Novoselov}, V.~S. {Beskin}, A.~K. {Galishnikova}, M.~M. {Rashkovetskyi}, and A.~V. {Biryukov}.
\newblock {Orthogonal pulsars as a key test for pulsar evolution}.
\newblock \emph{\mnras}, 494\penalty0 (3):\penalty0 3899--3911, April 2020.
\newblock \doi{2020MNRAS.494.3899N}.

\bibitem[{Philippov} and {Kramer}(2022)]{Philippov_Kramer}
A.~{Philippov} and M.~{Kramer}.
\newblock {Pulsar Magnetospheres and Their Radiation}.
\newblock \emph{\araa}, 60:\penalty0 495--558, August 2022.
\newblock \doi{10.1146/annurev-astro-052920-112338}.

\bibitem[{Philippov} et~al.(2014){Philippov}, {Tchekhovskoy}, and {Li}]{philippovtchekh2014}
A.~{Philippov}, A.~{Tchekhovskoy}, and J.~G. {Li}.
\newblock {Time evolution of pulsar obliquity angle from 3D simulations of magnetospheres}.
\newblock \emph{\mnras}, 441:\penalty0 1879--1887, July 2014.
\newblock \doi{10.1093/mnras/stu591}.

\bibitem[{Philippov} and {Spitkovsky}(2014)]{philippovspitkovsky2014}
A.~A. {Philippov} and A.~{Spitkovsky}.
\newblock {Ab Initio Pulsar Magnetosphere: Three-dimensional Particle-in-cell Simulations of Axisymmetric Pulsars}.
\newblock \emph{\apjl}, 785:\penalty0 L33, April 2014.
\newblock \doi{10.1088/2041-8205/785/2/L33}.

\bibitem[{Posselt} et~al.(2021){Posselt}, {Karastergiou}, {Johnston}, {Parthasarathy}, {Keith}, {Oswald}, {Song}, {Weltevrede}, {Barr}, {Buchner}, {Geyer}, {Kramer}, {Reardon}, {Serylak}, {Shannon}, {Spiewak}, and {Venkatraman Krishnan}]{MeerKAT_VI}
B.~{Posselt}, A.~{Karastergiou}, S.~{Johnston}, A.~{Parthasarathy}, M.~J. {Keith}, L.~S. {Oswald}, X.~{Song}, P.~{Weltevrede}, E.~D. {Barr}, S.~{Buchner}, M.~{Geyer}, M.~{Kramer}, D.~J. {Reardon}, M.~{Serylak}, R.~M. {Shannon}, R.~{Spiewak}, and V.~{Venkatraman Krishnan}.
\newblock {The Thousand-Pulsar-Array programme on MeerKAT - VI. Pulse widths of a large and diverse sample of radio pulsars}.
\newblock \emph{\mnras}, 508\penalty0 (3):\penalty0 4249--4268, December 2021.
\newblock \doi{10.1093/mnras/stab2775}.

\bibitem[{Rankin}(1983)]{Rankin1983}
J.~M. {Rankin}.
\newblock {Toward an empirical theory of pulsar emission. I. Morphological taxonomy.}
\newblock \emph{\apj}, 274:\penalty0 333--358, November 1983.
\newblock \doi{10.1086/161450}.

\bibitem[{Rankin}(1990)]{RankinIV}
J.~M. {Rankin}.
\newblock {Toward an empirical theory of pulsar emission. IV - Geometry of the core emission region}.
\newblock \emph{\apj}, 352:\penalty0 247--257, March 1990.
\newblock \doi{10.1086/168530}.

\bibitem[{Ruderman} and {Sutherland}(1975)]{RS}
M.~A. {Ruderman} and P.~G. {Sutherland}.
\newblock {Theory of pulsars - Polar caps, sparks, and coherent microwave radiation}.
\newblock \emph{\apj}, 196:\penalty0 51--72, February 1975.
\newblock \doi{10.1086/153393}.

\bibitem[{Spitkovsky}(2006)]{spitkovsky2006}
A.~{Spitkovsky}.
\newblock {Time-dependent Force-free Pulsar Magnetospheres: Axisymmetric and Oblique Rotators}.
\newblock \emph{\apjl}, 648:\penalty0 L51--L54, September 2006.
\newblock \doi{10.1086/507518}.

\bibitem[{Stinebring} et~al.(1984){Stinebring}, {Cordes}, {Rankin}, {Weisberg}, and {Boriakoff}]{Stinebring1984}
D.~R. {Stinebring}, J.~M. {Cordes}, J.~M. {Rankin}, J.~M. {Weisberg}, and V.~{Boriakoff}.
\newblock {Pulsar polarization fluctuations. I. 1404 MHz statistical summaries.}
\newblock \emph{\apjs}, 55:\penalty0 247--277, June 1984.
\newblock \doi{10.1086/190954}.

\bibitem[{Tauris} and {Manchester}(1998)]{T&M}
T.~M. {Tauris} and R.~N. {Manchester}.
\newblock {On the Evolution of Pulsar Beams}.
\newblock \emph{\mnras}, 298:\penalty0 625--636, August 1998.
\newblock \doi{10.1046/j.1365-8711.1998.01369.x}.

\bibitem[{Taylor} and {Manchester}(1975)]{Taylor_Manchester}
J.~H. {Taylor} and R.~N. {Manchester}.
\newblock {Observed properties of 147 pulsars.}
\newblock \emph{\aj}, 80:\penalty0 794--806, October 1975.
\newblock \doi{10.1086/111813}.

\bibitem[{Wang} et~al.(2010){Wang}, {Lai}, and {Han}]{wanglaihan2010}
C.~{Wang}, D.~{Lai}, and J.~{Han}.
\newblock {Polarization changes of pulsars due to wave propagation through magnetospheres}.
\newblock \emph{\mnras}, 403:\penalty0 569--588, April 2010.
\newblock \doi{10.1111/j.1365-2966.2009.16074.x}.

\bibitem[{Wang} et~al.(2023){Wang}, {Han}, {Xu}, {Wang}, {Yan}, {Jing}, {Su}, {Zhou}, and {Wang}]{FAST682}
P.~F. {Wang}, J.~L. {Han}, J.~{Xu}, C.~{Wang}, Y.~{Yan}, W.~C. {Jing}, W.~Q. {Su}, D.~J. {Zhou}, and T.~{Wang}.
\newblock {FAST Pulsar Database. I. Polarization Profiles of 682 Pulsars}.
\newblock \emph{Research in Astronomy and Astrophysics}, 23\penalty0 (10):\penalty0 104002, October 2023.
\newblock \doi{10.1088/1674-4527/acea1f}.

\bibitem[{Weltevrede} and {Johnston}(2008{\natexlab{a}})]{WJ2008}
P.~{Weltevrede} and S.~{Johnston}.
\newblock {Profile and polarization characteristics of energetic pulsars}.
\newblock \emph{\mnras}, 391:\penalty0 1210--1226, December 2008{\natexlab{a}}.
\newblock \doi{10.1111/j.1365-2966.2008.13950.x}.

\bibitem[{Weltevrede} and {Johnston}(2008{\natexlab{b}})]{WJ2008b}
P.~{Weltevrede} and S.~{Johnston}.
\newblock {The population of pulsars with interpulses and the implications for beam evolution}.
\newblock \emph{\mnras}, 387:\penalty0 1755--1760, July 2008{\natexlab{b}}.
\newblock \doi{10.1111/j.1365-2966.2008.13382.x}.

\bibitem[{Young} et~al.(2010){Young}, {Chan}, {Burman}, and {Blair}]{Young2010}
M.~D.~T. {Young}, L.~S. {Chan}, R.~R. {Burman}, and D.~G. {Blair}.
\newblock {Pulsar magnetic alignment and the pulsewidth-age relation}.
\newblock \emph{\mnras}, 402\penalty0 (2):\penalty0 1317--1329, February 2010.
\newblock \doi{10.1111/j.1365-2966.2009.15972.x}.

\bibitem[Торопов et~al.(2024)Торопов, Тюльбашев, and Бескин]{Toporov}
М.~О. Торопов, С.~А. Тюльбашев, and В.~С. Бескин.
\newblock Поиск интерпульсов в полной выборке пульсаров на частоте 111 МГц.
\newblock \emph{Астрономический журнал}, 101\penalty0 (12):\penalty0 1084--1094, 2024.
\newblock ISSN 0004-6299.
\newblock URL \url{https://journals.rcsi.science/0004-6299/article/view/276064}.

\end{thebibliography}

\end{document}